\documentclass[conference]{IEEEtran}

\usepackage[T1]{fontenc}
\usepackage{color, colortbl}
\usepackage{babel}
\usepackage{verbatim}
\usepackage{textcomp}
\usepackage{amsmath}
\usepackage{amsthm}
\usepackage{amssymb}
\usepackage{amstext}
\usepackage{graphicx}
\usepackage{tabulary}
\usepackage{caption}
\usepackage{subcaption}
\usepackage{epstopdf}
\usepackage{cite}
\usepackage[normalem]{ulem}
\usepackage{balance}
\usepackage{multirow}
\usepackage{xcolor}
\usepackage{soul}
\usepackage[ruled,vlined]{algorithm2e}
\addto\captionsenglish{}

\def\RSS{\mathtt{RSS}}

\newlength \figwidth
\newcommand{\green}[1]{{\textcolor[rgb]{0,0.5,0}{#1}}}

\newcommand{\gio}[1]{\noindent \green{ {{$\blacktriangleright$ 
   {\textsf{[Gio]: #1}} $\blacktriangleleft$}}}}

\newtheorem{Proposition}{\bf Proposition}

\newtheorem{Remark}{\bf Remark}

\makeatletter

 \let\oldforeign@language\foreign@language
 \DeclareRobustCommand{\foreign@language}[1]{%
   \lowercase{\oldforeign@language{#1}}}

\def\nb0{{\mathbf{0}}}
\def\nb1{{\mathbf{1}}}

\makeatother

\IEEEoverridecommandlockouts

\begin{document}
\bstctlcite{IEEEexample:BSTcontrol}

\title{Analysis of UAV Corridors in Cellular Networks}

\author{\IEEEauthorblockN{{Saeed~Karimi-Bidhendi$^{\star}$, Giovanni~Geraci$^{\sharp}$, and Hamid~Jafarkhani$^{\star}$}} \vspace{0.2cm}
\IEEEauthorblockA{$^{\star}$\emph{Center for Pervasive
Communications and Computing, Univ. of California, Irvine CA, USA}\\
\IEEEauthorblockA{$^{\sharp}$\emph{Universitat Pompeu Fabra (UPF), Barcelona, Spain}
}}
\thanks{This work was supported in part by the NSF Award CNS-2229467 and by the Spanish State Research Agency through grants RTI2018-101040-A-I00, PID2021-123999OB-I00, and through the ``Ram\'{o}n y Cajal'' program.}
}

\maketitle

\begin{abstract}
In this article, we introduce a new mathematical framework for the analysis and design of UAV corridors in cellular networks, while considering a realistic network deployment, antenna radiation pattern, and propagation channel model. By leveraging quantization theory, we optimize the electrical tilts of existing ground cellular base stations to maximize the coverage of both legacy ground users and UAVs flying along specified aerial routes. Our practical case study shows that the optimized network results in a cell partitioning that significantly differs from the usual hexagonal pattern, and that it can successfully guarantee coverage all over the UAV corridors without degrading the perceived signal strength on the ground.
\end{abstract}

\section{Introduction}




Barely seen in action movies until a decade ago, the progressive blending of uncrewed aerial vehicles (UAVs) into our daily lives will enhance safety and greatly impact labor and leisure activities alike. Most stakeholders regard reliable connectivity as a must-have for the UAV ecosystem to thrive. As a result, UAV cellular communications have witnessed a surge of interest in terms of (i) what networks can do for UAVs and (ii) what UAVs can do for networks \cite{ZenGuvZha2020,SaaBenMoz2020,NamChaKim17,wu20205g,CDHJGC20}.


As for (i)---focus of the present paper---the mobile industry and its academic research counterpart have long joined forces to pursue reliable connectivity up in the air by re-engineering existing terrestrial networks, originally designed for ground users only \cite{3GPP36777,GerGarAza2022}.
Recent ideas for ubiquitous aerial connectivity hinge, e.g., on network densification \cite{GarGerLop2019,GerGarGal2018,KanMezLoz2021,DanGarGer2020,CDALHJTWC22}, dedicated infrastructure for aerial services \cite{GerLopBen2022,MozLinHay2021}, or leveraging satellites to complement the ground network \cite{BenGerLop2022}, all requiring costly hardware or signal processing upgrades.


Fortunately, many impactful UAV use cases could still be enabled by providing reliable connectivity along predetermined aerial routes, i.e., \emph{UAV corridors}, enforced by the appropriate traffic authorities \cite{CheJaaYan2020,BhuGuvDai2021}. The research community has started contributing in this direction by studying UAV trajectory optimization, e.g., matching the route of a UAV to the best coverage pattern provided by the network  \cite{BulGuv2018,ChaSaaBet2018,EsrGanGes2020,BayTheCac2021}.
More recent work has targeted tuning cellular deployments to cater for UAV corridors through system-level simulations, large-scale optimization, or the theoretical analysis of a simplified setup \cite{MaeChoGuv2021,ChoGuvSaa2021,SinBhaOzt2021,BerLopGes2022}. 
However, there is an unmet need for a general mathematical framework allowing the analysis and design of UAV corridors in cellular networks.

%

In this paper, we take the first step towards creating such mathematical framework through quantization theory  \cite{GuoJaf2016,guo2018source,guo2019movement,karimi2020energy,karimi2021energy,9086619,8519749}. Specifically, we determine the necessary conditions and design an iterative algorithm to optimize the antenna tilts at each base station of a cellular network for a maximum received signal strength (RSS)---a proxy for coverage---at both legacy ground users and UAVs flying along corridors. To the best of our knowledge, this is the first work doing so in a rigorous yet tractable manner, while accounting for a realistic network deployment, antenna radiation pattern, and propagation channel model.
We further put our mathematical framework into practice with a case study, whose main takeaways can be summarized as follows:
\begin{itemize}
    \item Pursuing satisfactory coverage of both ground users and UAV corridors can result in a highly uncustumary cell partitioning that profoundly differs from a hexagonal pattern, commonplace in legacy ground-only systems.
    \item By tilting up a selected subset of base stations, one can significantly boost the RSS along multiple UAV corridors compared to an all-downtilt baseline, achieving levels of aerial coverage close to an upper bound arrangement that ignores legacy ground users altogether.
    \item Additionally, one can nearly preserve the quality coverage on the ground, maintaining levels of RSS close to those experienced in a scenario devoid of UAVs.
\end{itemize}

\begin{figure}
\centering
\includegraphics[width=0.48\textwidth]{
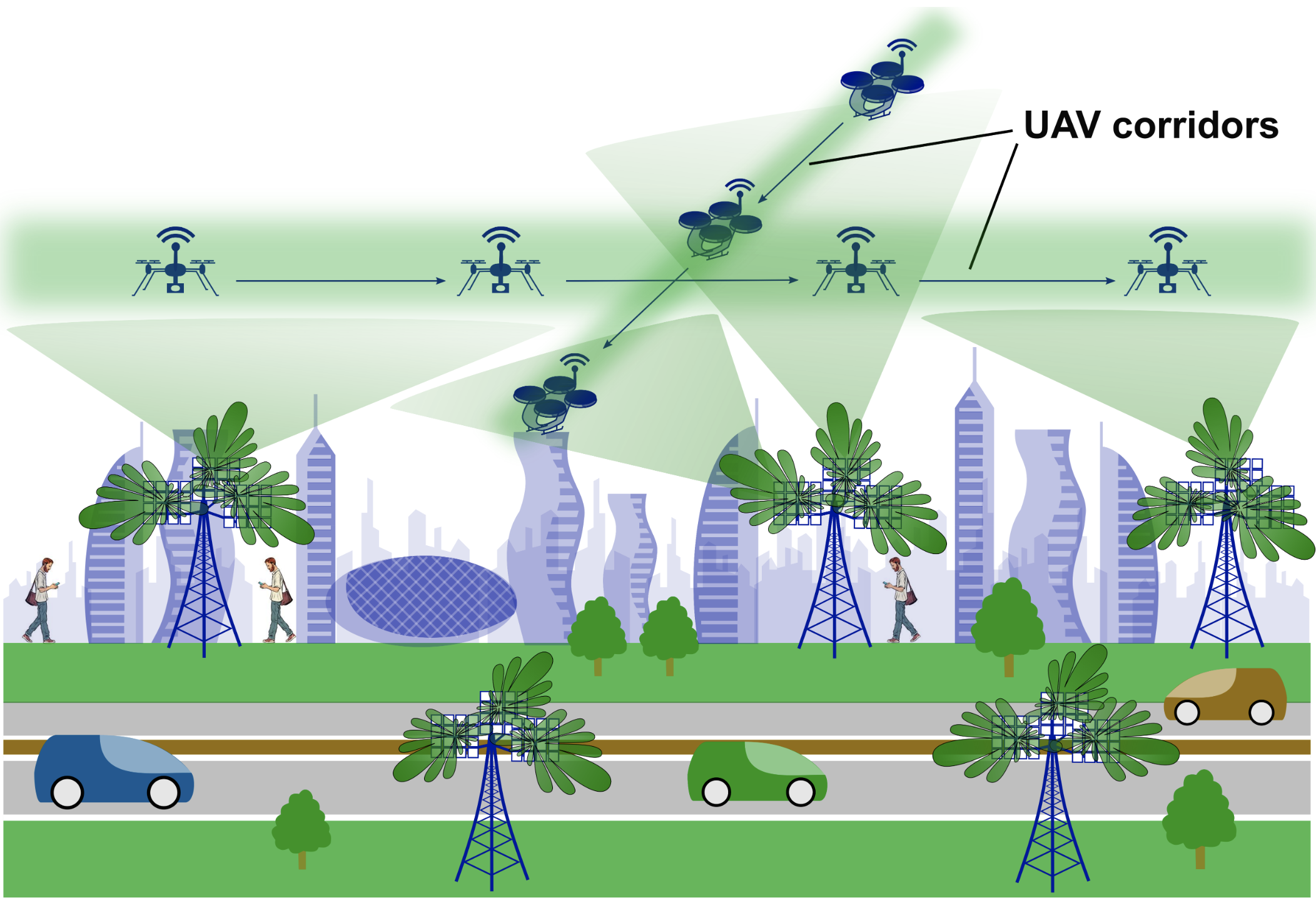}
\caption{Illustration of a cellular network with downtilted and uptilted base stations providing coverage to ground users as well as UAVs flying along corridors (blurred green).}
\label{fig:illustration}
\end{figure}

\section{System Model}\label{System_Model}

The set-up under consideration is illustrated in Fig.~\ref{fig:illustration} and detailed as follows.


\subsection{Network Topology}\label{Network_Topology}

\subsubsection{Ground cellular network}
The backbone of our network is 
a ground cellular deployment comprising $N$ 
base stations (BSs) that provide coverage and service for network users. For each $i\in \{1,\cdots, N\}$, we denote the height and 2D location of BS $i$ by $h_{i,\mathrm{B}}$ and $p_i$, respectively. 
Let $\theta_i \in \left[ -90^{\circ}, +90^{\circ}\right]$ be the vertical antenna tilt of BS $i$, which can be optimized by a mobile operator, with positive and negative angles denoting uptilts and downtilts, respectively.
Let $\phi_i \in \left[ -180^{\circ}, +180^{\circ}\right]$ be the antenna horizontal boresight direction (azimuth) of BS $i$, assumed fixed given the deployment.

\subsubsection{UAV corridors and legacy ground users} Our network entails two types of users: UAVs and ground-users (GUEs). 
UAVs traverse a 2D region $Q_{\mathrm{U}} = \bigcup_{u=1}^{N_U} Q_{\mathrm{u}}$ consisting of $N_U$ UAV corridors $Q_{\mathrm{u}}$, i.e., predefined 2D aerial regions. For each corridor $Q_{\mathrm{u}}$, all UAVs are assumed to fly at the same height $h_{\mathrm{u}}$. Ground users (GUEs) populate a 2D region $Q_{\mathrm{G}}$ 
and have a fixed height $h_{\mathrm{G}}$.
Let $\lambda(q)$  be a probability density function that reflects the distribution of users in  $Q= Q_{\mathrm{U}} \cup Q_{\mathrm{G}}$. Each user in $Q$ is associated with the BS providing the largest received signal strength (RSS), defined in the sequel. Hence, the region $Q$ can be partitioned into $N$ different subregions $\mathbf{V}=(V_1,\cdots,V_N)$ such that users in $V_i$ are associated with BS $i$.

\subsection{Channel Model and Performance Metric}\label{Channel_Model_and_Performance_Metric}

\subsubsection{Antenna gain}
We assume BSs equipped with directional antennas. The vertical and horizontal half-power beamwidths of the directional antennas are denoted by $\theta_{\text{3dB}}$ and $\phi_{\text{3dB}}$, respectively. The total antenna gain of BS $i$ in dB is given by
\begin{equation}
A_{i,q} = A_{\textrm{max}} + A_{i,q}^{\mathrm{V}} + A_{i,q}^{\mathrm{H}},
\end{equation}
where $A_{\textrm{max}}$ denotes the maximum antenna gain at the boresight, $A_{i,q}^{\mathrm{V}}$ and $A_{i,q}^{\mathrm{H}}$ denote the vertical and horizontal antenna gains in dB, respectively, given by \cite{3GPP38901}
\begin{align}\label{eqn:antenna_V}
    A_{i,q}^{\mathrm{V}} &=
    - \frac{12}{\theta^2_{\text{3dB}}} \left[ \theta_{i,q} - \theta_i \right]^2, \\    
    A_{i,q}^{\mathrm{H}} &=
    - \frac{12}{\phi^2_{\text{3dB}}} \left[ \phi_{i,q} - \phi_i \right]^2.
\label{eqn:antenna_H}
\end{align}

The vertical antenna gain $A_{i,q}^{\mathrm{V}}$ in (\ref{eqn:antenna_V}) depends on the vertical tilt $\theta_i$ of BS $i$ and on the elevation angle $\theta_{i,q}$ between BS $i$ and user location $q\in Q$, calculated as 
\begin{equation}
    \theta_{i,q} = 
    \tan^{-1}\left( \frac{h_q - h_{i,\mathrm{B}}}{\| q - p_i \|}  \right),
\label{eqn:theta_iq}
\end{equation}
where $\|\cdot\|$ denotes $l^2$ norm. The horizontal antenna gain $A_{i,q}^{\mathrm{H}}$ in (\ref{eqn:antenna_H}) depends on the difference $\left[\phi_{i,q} - \phi_i\right] \in \left[ -180^{\circ}, +180^{\circ}\right]$, where $\phi_i$ is the azimuth orientation of BS $i$ and $\phi_{i,q}$ is the azimuth angle between BS $i$ and user location $q$, given by 
\begin{equation}
\phi_{i,q} \!=\! 
\begin{cases}
\!\tan^{-1}\!\!\left(\frac{q_{\mathrm{y}}-p_{i,\mathrm{y}}}{q_{\mathrm{x}}-p_{i,\mathrm{x}}}\right) \!+\! 180^{\circ}\!\times\! 2c &  \text{if $q_{\mathrm{x}}-p_{i,\mathrm{x}}>0$}\\
\!\tan^{-1}\!\!\left(\frac{q_{\mathrm{y}}-p_{i,\mathrm{y}}}{q_{\mathrm{x}}-p_{i,\mathrm{x}}}\right) \!+\! 180^{\circ}\!\times\! (2c \!+\! 1) &  \text{if $q_{\mathrm{x}}-p_{i,\mathrm{x}}<0$}
\end{cases}
\label{eqn:phi_iq}
\end{equation}
where subscripts $\cdot_{\mathrm{x}}$ and $\cdot_{\mathrm{y}}$ denote the horizontal and vertical coordinates of a point, respectively, and the integer $c$ is chosen in a way that $-180^{\circ} \leq \phi_{i,q} - \phi_i \leq +180^{\circ}$. 


\subsubsection{Pathloss}
The distance-dependent pathloss between a user at location $q$ and BS $i$ is given by
\begin{equation} 
L_{i,q} = a_q + b_q~\log_{10}\left[\| q - p_i \|^2 + (h_q-h_{i,\mathrm{B}})^2 \right]^{\frac{1}{2}},
\label{eqn:Pathloss}
\end{equation} 
where for each user location $q$, the constants $a_q$ and $b_q$ depend on the carrier frequency and on the pathloss exponent, the latter affected by the BS deployment features, the height of the user at $q$ and the link's line-of-sight (LoS) condition. In our numerical simulations, we employ practical values for the parameters $a_q$ and $b_q$, obtained from \cite{3GPP36777,3GPP38901} and reported in Section \ref{Case_Study_Channel}.

\subsubsection{Received signal strength}\label{Received_signal_strength}
The RSS in dBm from BS $i$ at user location $q$ is given by\footnote{While our analysis is general and can incorporate shadow fading, we here neglect it to focus on the role played by optimizing the vertical tilts $\theta_i$.}
\begin{equation}
\begin{aligned}
&\RSS^{(i)} (q; \theta_i)= \rho_i + A_{i,q} - L_{i,q} \quad \text{[dBm]}\\
&\quad = \rho_i + A_{\textrm{max}} - \frac{12}{\theta^2_{\text{3dB}}} \left[ \theta_{i,q} - \theta_i \right]^2 - \frac{12}{\phi^2_{\text{3dB}}} \left[ \phi_{i,q} - \phi_i \right]^2 \\
&\quad - a_q - b_q~\log_{10}\left[\| q - p_i \|^2 + (h_q-h_{i,\mathrm{B}})^2 \right]^{\frac{1}{2}},  \text{ [dBm]}
\end{aligned}
\label{eqn:RSS}
\end{equation}
where $\rho_i$ denotes BS $i$'s transmit power  measured in dBm.

\subsection{Performance Function}\label{Performance_Function}

In the remainder of the paper, we assume that all BS locations $p_i$ and azimuth orientations $\phi_i$ are fixed while all BS vertical antenna tilts $\theta_i$ are optimized. 
Our main goal is to maximize the provided received signal strength averaged over all users in the network. 

\begin{Remark}
Optimizing the vertical tilts $\theta_i$ entails that each BS has a different value of $\theta_i$ and therefore a different received signal strength 
$\RSS^{(i)} (q; \theta_i)$. Moreover, the latter is not necessarily a non-increasing function of the distance $\| q - p_i \|$, e.g., moving away from a BS can sometimes yield a worse pathloss $L_{i,q}$ but a better antenna gain $A_{i,q}$.
\end{Remark}

The overall performance function, i.e., RSS in dBm averaged over all users in the network, is given by:
\begin{align}
    \Phi(\mathbf{V}, \mathbf{\Theta}) &=  \sum_{i=1}^{N} \int_{V_i} \RSS^{(i)} (q; \theta_i) \lambda(q) dq.
\label{eqn:Phi}
\end{align}
Our goal is to optimize the performance function $\Phi(\mathbf{V}, \mathbf{\Theta})$ over the cell partitioning $\mathbf{V}=(V_1,\cdots,V_N)$ and BS vertical antenna tilts $\mathbf{\Theta} = (\theta_1, \cdots, \theta_N)$. Note that, while other choices for $\Phi(\mathbf{V}, \mathbf{\Theta})$ are not precluded, averaging the RSS in dBm---i.e., in logarithmic scale and thus equivalent to a max-product criterion in linear units---pursues fairness among users, as it will be shown in our case study.

\section{Analytical Framework}

As shown in (\ref{eqn:Phi}), the performance function $\Phi$ depends on both variables $\mathbf{V}$ and $\mathbf{\Theta}$; thus, our goal is to find the optimal cell partitioning $\mathbf{V}^*=\left(V_1^*, \cdots, V_N^* \right)$ and vertical antenna tilts $\mathbf{\Theta}^* = (\theta_1^*, \cdots, \theta_N^*)$ that maximize the performance function. Note that not only the variables $\mathbf{V}$ and $\mathbf{\Theta}$ are interdependent, i.e., the optimal value for each variable depends on the value of the other variable, but also this is an NP-hard optimization problem. Our aim is to develop an alternating optimization algorithm that iteratively updates the values of $\mathbf{V}$ and $\mathbf{\Theta}$. We accomplish this goal in two optimal steps: (i) updating the cell partitioning $\mathbf{V}$ for a given set of vertical antenna tilts $\mathbf{\Theta}$; and (ii) updating the BS vertical antenna tilts $\mathbf{\Theta}$ for a given cell partitioning $\mathbf{V}$. The following proposition provides the necessary condition and update rule for Step (i):
\begin{Proposition}\label{optimal_cell_partitioning}
The optimal cell partitioning $\mathbf{V}^*$ for a given set of vertical antenna tilts $\mathbf{\Theta}$ is given by:
\begin{align}\label{updated_eq2}
        V_n^*(\mathbf{\Theta}) = \left\{ q \in Q  |  \RSS^{(n)} (q; \theta_n)
      \geq \RSS^{(k)} (q; \theta_k) , \forall k \in 1,\ldots,
     N\right\},
\end{align}
for each $n\in\{1,\cdots,N\}$.
\end{Proposition}
\textit{Proof. }Given any arbitrary cell partitioning of $Q$ such as  $\mathbf{W}=(W_1,\cdots,W_N)$, we have:
\begin{align}
    \Phi(\mathbf{W},\mathbf{\Theta}) &= \sum_{n=1}^{N} \int_{W_n} \RSS^{(n)} (q; \theta_n) \lambda(q) dq \\
    &\leq \sum_{n=1}^{N} \int_{W_n} \max_k \Big[\RSS^{(k)} (q; \theta_k)\Big] \lambda(q) dq \\
    &=\int_Q \max_k \Big[\RSS^{(k)} (q; \theta_k)\Big] \lambda(q) dq \\
    &=\sum_{n=1}^{N} \int_{V_n^*} \max_k \Big[\RSS^{(k)} (q; \theta_k)\Big] \lambda(q) dq \\
    &=\sum_{n=1}^{N} \int_{V_n^*} \RSS^{(n)} (q; \theta_n) \lambda(q) dq \\
    &= \Phi(\mathbf{V}^*, \mathbf{\Theta}).
\end{align}
Thus, $\mathbf{V}^*$ yields the maximum performance and is optimal.$\hfill\blacksquare$

Now, for the second step, we aim to find the optimal $\mathbf{\Theta}^*$ that maximizes the performance function $\Phi$ for a given cell partitioning $\mathbf{V}$. Our approach is to apply the gradient ascent algorithm to find the optimal vertical antenna tilts. Gradient ascent is a first-order iterative optimization algorithm for finding a local maximum of a differentiable function. The idea is to take repeated scaled steps in the direction of the gradient since this is the direction of steepest ascent.

\begin{Proposition}\label{gradient_vector}
For each $n\in\{1,\cdots,N\}$, the partial derivative of the performance function in (\ref{eqn:Phi}) w.r.t. $\theta_n$ is given by 
\begin{align}
\frac{\partial \Phi(\mathbf{V},\mathbf{\Theta})}{\partial \theta_n}  =   \frac{24}{\theta^2_{\text{3dB}}} \Bigg\{ \sum_{u=1}^{N_U}
     & \int_{V_n(\mathbf{\Theta})\cap Q_u} \!\!\!\!\!\!\!\! (\theta_{n,q}-\theta_n) \lambda(q) dq  \nonumber\\ + & \int_{V_n(\mathbf{\Theta})\cap Q_G} \!\!\!\!\!\!\!\! (\theta_{n,q}-\theta_n) \lambda(q) dq \Bigg\}.
\label{eqn:derivativePhidBm}
\end{align}
\end{Proposition}
\textit{Proof. }
The derivative of (\ref{eqn:Phi}) contains two terms: (i) the derivative of the integrand, and (ii) the integral over the boundaries. According to (\ref{updated_eq2}), for a point $q$ on the boundary of regions $n$ and $k$, we have: $\RSS^{(n)}(q; \theta_n) = \RSS^{(k)}(q; \theta_k)$. Since the normal outward vectors at point $q$ have opposite directions for these two regions, the term (ii) amounts to zero \cite{GuoJaf2016}. The gradient $\frac{\partial \Phi(\mathbf{\Theta})}{\partial \theta_n}$ is then given by the first term, obtained as:
\begin{align}
    \frac{\partial \Phi(\mathbf{V},\mathbf{\Theta})}{\partial \theta_n} = \int_{V_n(\mathbf{\Theta})} \frac{\partial}{\partial \theta_n} \RSS^{(n)}(q; \theta_n) \lambda(q)dq \nonumber \\
    \stackrel{(\text{a})}{=}  \frac{24}{\theta^2_{\text{3dB}}} \Bigg\{ \sum_{u=1}^{N_U}
      \int_{V_n(\mathbf{\Theta})\cap Q_u} \!\!\!\!\!\!\!\! (\theta_{n,q}-\theta_n) \lambda(q) dq  \nonumber\\ +  \int_{V_n(\mathbf{\Theta})\cap Q_G} \!\!\!\!\!\!\!\! (\theta_{n,q}-\theta_n) \lambda(q) dq \Bigg\}.
    \label{eqn:derivativePhi}
\end{align}
where (a) follows from the definition of $Q$.  $\hfill\blacksquare$

Propositions \ref{optimal_cell_partitioning} and \ref{gradient_vector} provide the main ingredients to design the BS vertical antenna tilt (BS-VAT) optimization algorithm outlined in Algorithm \ref{BS_VAT_Algorithm}.


\begin{algorithm}[ht!]
\SetAlgoLined
\SetKwRepeat{Do}{do}{while}
\KwResult{Optimal BS vertical antenna tilts $\mathbf{\Theta}^*$ and cell partitioning $\mathbf{V}^*$.}
\textbf{Input:} Initial BS vertical antenna tilts $\mathbf{\Theta}$ and cell partitioning $\mathbf{V}$, learning rate $\eta_0\in (0,1)$, convergence error thresholds $\epsilon_1, \epsilon_2\in \mathbb{R}^+$, constant $\kappa \in (0, 1)$ \;

 \Do{$\frac{\Phi_{\textrm{new}} - \Phi_{\textrm{old}}}{\Phi_{\textrm{old}}} \geq \epsilon_2$}
 {
Calculate  $\Phi_{\textrm{old}} = \Phi\left(\mathbf{V},\mathbf{\Theta}\right)$\;
{\color{gray}\# {\it Update the cell partitioning $\mathbf{V}$};}\\
\For{ $n\in \{1,\cdots,N\}$} {
        Update the cell $V_n$ according to Eq. (\ref{updated_eq2});
}
{\color{gray}\# {\it Start the gradient ascent algorithm};}\\    
Set $\eta \gets \eta_0$\;
\Do{$\frac{\Phi_{\textrm{e}} - \Phi_{\textrm{s}}}{\Phi_{\textrm{s}}} \geq \epsilon_1$}
{
Calculate  $\Phi_{\textrm{s}} = \Phi\left(\mathbf{V},\mathbf{\Theta}\right)$\;
{\color{gray}\# {\it Calculate the gradient $\nabla_{\mathbf{\Theta}} \Phi(\mathbf{V},\mathbf{\Theta})$};}\\
\For{ $n\in \{1,\cdots,N\}$} {
        Calculate $\frac{\partial \Phi(\mathbf{V},\mathbf{\Theta})}{\partial \theta_n}$ according to Eq. (\ref{eqn:derivativePhi});
}
{\color{gray}\# {\it Update the learning rate};}\\
$\eta \gets \eta \times \kappa$\;
{\color{gray}\# {\it Update the vertical antenna tilts $\mathbf{\Theta}$};}\\    
$\mathbf{\Theta} \gets \mathbf{\Theta} + \eta \nabla_{\mathbf{\Theta}} \Phi(\mathbf{V},\mathbf{\Theta})$\;
Calculate $\Phi_{\textrm{e}} = \Phi\left(\mathbf{V},\mathbf{\Theta}\right)$\;
}
Calculate  $\Phi_{\textrm{new}} = \Phi\left(\mathbf{V},\mathbf{\Theta}\right)$\;
}
 \caption{BS vertical antenna tilt optimization}
 \label{BS_VAT_Algorithm}
\end{algorithm}

\begin{Proposition}\label{convergence_theorem}
The BS-VAT algorithm is an iterative improvement algorithm and  converges. 
\end{Proposition}
\textit{Proof. }We demonstrate that none of the two steps in the BS-VAT algorithm decreases the performance function $\Phi$ in (\ref{eqn:Phi}). In the first step, the cell partitioning $\mathbf{V}$ is updated according to (\ref{updated_eq2}) while $\mathbf{\Theta}$ is fixed. Proposition \ref{optimal_cell_partitioning} indicates that the obtained $\mathbf{V}$ is optimal for the current set of antenna tilts $\mathbf{\Theta}$. Thus, the first step does not decrease the performance function $\Phi$. In the second step, the gradient ascent algorithm is utilized to optimize $\mathbf{\Theta}$ while $\mathbf{V}$ is fixed. Note that the learning rate at iteration $t$ of the gradient ascent algorithm is equal to $\eta_t = \eta_0\times \kappa^t$. Since $\sum_{t=1}^{\infty}\eta^2_t < \sum_{t=1}^{\infty}\eta_t = \frac{\kappa}{1 - \kappa}\eta_0 < \infty$, the gradient ascent is guaranteed to converge \cite{goodfellow2016deep} and does not decrease the performance function $\Phi$. Hence, the BS-VAT algorithm generates a sequence of non-decreasing performance function values. Since the performance function $\Phi(\mathbf{V}, \mathbf{\Theta})$ is also upper bounded because of the limited transmission power at each base station, 
the algorithm converges.  $\hfill\blacksquare$

\section{Case Study}\label{Case_Study}

To evaluate the performance of our theoretical framework, we consider a case study in this section. 

\begin{table}
\centering
\vspace{1mm}
\caption{System-level parameters for our case study.}
\label{table:parameters}
\def\arraystretch{1.2}
\begin{tabulary}{\columnwidth}{ |p{1.7cm} | p{6.1cm} | }
\hline
  \textbf{Deployment}&  \\ \hline
  $P$		           & Hexagonal grid with intersite distance $\mathrm{ISD} = 500$~m. Two tiers of BSs around the one at the origin, three sectors per site, 57 BSs in total, $h_{\mathrm{B}}=25$~m. \\ \hline
  $Q_{\mathrm{U}}$   & Consisting of $N_U=4$ aerial corridors \\\hline
  $Q_u, u=1$		   & Vertical $[-320, -280] \times [-400, 400]$, $h_{\mathrm{u}}$ = $150$~m \\\hline
  $Q_u, u=2$		   & Vertical $[-120, -80] \times [-400, 400]$, $h_{\mathrm{u}}$ = $120$~m \\\hline
  $Q_u, u=3$	       & Vertical $[80, 120]\times [-400, 400]$, $h_{\mathrm{u}}$ = $120$~m \\\hline
  $Q_u, u=4$		   & Vertical $[280, 320]\times [-400, 400]$, $h_{\mathrm{u}}$ = $150$~m  \\\hline
  $Q_{\mathrm{G}}$   & Square area $[-750, 750] \times [-750, 750]$, $h_{\mathrm{G}}$ = $1.5$~m \\ \hline
 $\lambda_\textrm{G}(q)$, $\lambda_\textrm{U}(q)$       & Uniform in $Q_\textrm{G}$ and $Q_\textrm{U}$, respectively \\ \hline
  $\lambda(q)$       & $\alpha\lambda_\textrm{G}(q) + (1-\alpha) \lambda_\textrm{U}(q)$ with $\alpha=\{1, 0, 0.5\}$ \\ \hline\hline
  \textbf{Channel}	                       &  \\ \hline
  $A_{\text{max}}$, $\rho_i$                 & 14~dBi, 43~dBm $\forall i$, respectively \\ \hline
  $\theta_{\text{3dB}}$, $\phi_{\text{3dB}}$ & $10^{\circ}$, $65^{\circ}$, respectively \\ \hline
  $\phi_i$                                   & Fixed for the three sectors: $\phi_i \in \left\{ 0^{\circ}, 120^{\circ}, 240^{\circ} \right\}$ \\ \hline
  \multirow{2}{*}{$a_q$}
                        & $q\in Q_\textrm{U}$: 34.02~dB (carrier at 2~GHz)\\ \cline{2-2}
                        & $q\in Q_\textrm{G}$: 38.42~dB (carrier at 2~GHz)\\ \hline
  \multirow{2}{*}{$b_q$}
                        & $q\in Q_\textrm{U}$: 22 (i.e., pathloss exponent 2.2)\\ \cline{2-2} 
                        & $q\in Q_\textrm{G}$: 30 (i.e., pathloss exponent 3.0)\\ \hline
  \hline
  \textbf{Optimization}	&   \\ \hline
  Initial tilts & $\theta_i=0$ $\forall i$ \\ \hline
  Initial partition & Each $q\in Q$ assigned to a random base station \\ \hline
  $\eta_0$, $\kappa$, $\epsilon_1$, $\epsilon_2$ & $0.005$, $0.999$, $10^{-8}$, $10^{-9}$, respectively \\ \hline
\end{tabulary}
\vspace{-0.4cm}
\end{table}


\subsection{Deployment Setup}\label{Case_Study_Deployment}

Simulations are carried out for a practical cellular network consisting of $19$ sites. Each site, say $i$, includes three sectors, i.e., three cells with the corresponding BSs placed at the exact same locations $p_{3\times i - 2} = p_{3\times i - 1} = p_{3\times i}$ but with different azimuth orientations $\phi_{3\times i - 2} = 0^{\circ}$, $\phi_{3\times i - 1} = 120^{\circ}$, and $\phi_{3\times i} = 240^{\circ}$. Thus, overall, there are $N=57$ BSs with corresponding vertical antenna tilts to optimize. The BSs are placed on a hexagonal layout with inter-site distance $\textrm{ISD} = 500\textrm{m}$ as illustrated in Fig. \ref{Fig:cellPartitioning}. Site indices are provided in Fig. \ref{Fig:cellPartitioningUAV}. All BSs are assumed to have the same height and transmission power, with $h_\textrm{i,B}=25$m and $\rho_i = 43$dBm $\forall i$, respectively. Ground users are distributed over a square area $Q_G = [-750, 750] \times [-750, 750]$ according to a uniform density function $\lambda_\textrm{G}(q)$ and are assumed to have the fixed height $h_\textrm{G} = 1.5$m. UAVs are distributed over $N_U = 4$ vertical aerial corridors, i.e., $Q_U = Q_1\cup Q_2\cup Q_3 \cup Q_4$, according to a uniform density function $\lambda_\textrm{U}(q)$. These corridors, shown in Fig.~\ref{Fig:cellPartitioningUAV}, are located at $Q_1 = [-320, -280]\times [-400, 400]$, $Q_2 = [-120, -80]\times [-400, 400]$, $Q_3 = [80, 120]\times [-400, 400]$, and $Q_4 = [280, 320]\times [-400, 400]$ and their heights are set at $h_1 = h_4 = 150$m and $h_2 = h_3 = 120$m. The density function $\lambda(q)$, which represents the distribution of users in $Q = Q_G \cup Q_U$, is a mixture of $\lambda_\textrm{G}(q)$ and $\lambda_\textrm{U}(q)$, i.e., $\lambda(q) = \alpha\lambda_\textrm{G}(q) + (1-\alpha) \lambda_\textrm{U}(q)$ where $\alpha$ is the mixing ratio. In the sequel, we consider three values for the parameter $\alpha$, namely $1$, $0$, and $0.5$. These three values correspond to optimizing the cellular network for ground users only, for UAVs only, and for both, respectively. 

\subsection{Channel Setup}\label{Case_Study_Channel}

As per 3GPP specifications \cite{3GPP36777,3GPP38901}, for a carrier frequency at $2$GHz and LoS condition, the values of $a_q$ and $b_q$ are set as:

\begin{equation}\label{a_q_values}
a_q = 
    \begin{cases}
      34.02 \textrm{ dB}, & \text{if}\ q\in Q_U, \\
      38.42 \textrm{ dB}, & \text{if}\ q\in Q_G,
    \end{cases}
\end{equation}
\begin{equation}\label{b_q_values}
b_q = 
    \begin{cases}
      22 \textrm{ (i.e., pathloss exponent 2.2)}, & \text{if}\ q\in Q_U, \\
      30 \textrm{ (i.e., pathloss exponent 3.0)}, & \text{if}\ q\in Q_G.
    \end{cases}
\end{equation}
The vertical and horizontal half-power beamwidth of the directional antennas are set to $\theta_\textrm{3dB} = 10^{\circ}$ and $\phi_\textrm{3dB} = 65^{\circ}$, respectively. The maximum antenna gain at the boresight is set to $A_\textrm{max} = 14$dBi.

\subsection{Vertical Antenna Tilt Optimization 
}\label{Case_Study_HPBW10}

The BS-VAT algorithm is initialized by a random cell partitioning, i.e., randomly assigning each $q\in Q$ to a base station, and setting $\theta_i = 0$ $\forall i$. The learning rate $\eta_0$ and constant $\kappa$ are set to $0.005$ and $0.999$, respectively. The convergence error thresholds are set to $\epsilon_1 = 10^{-8}$ and $\epsilon_2 = 10^{-9}$.

\begin{figure}
\centering
\includegraphics[width=0.49\textwidth]{
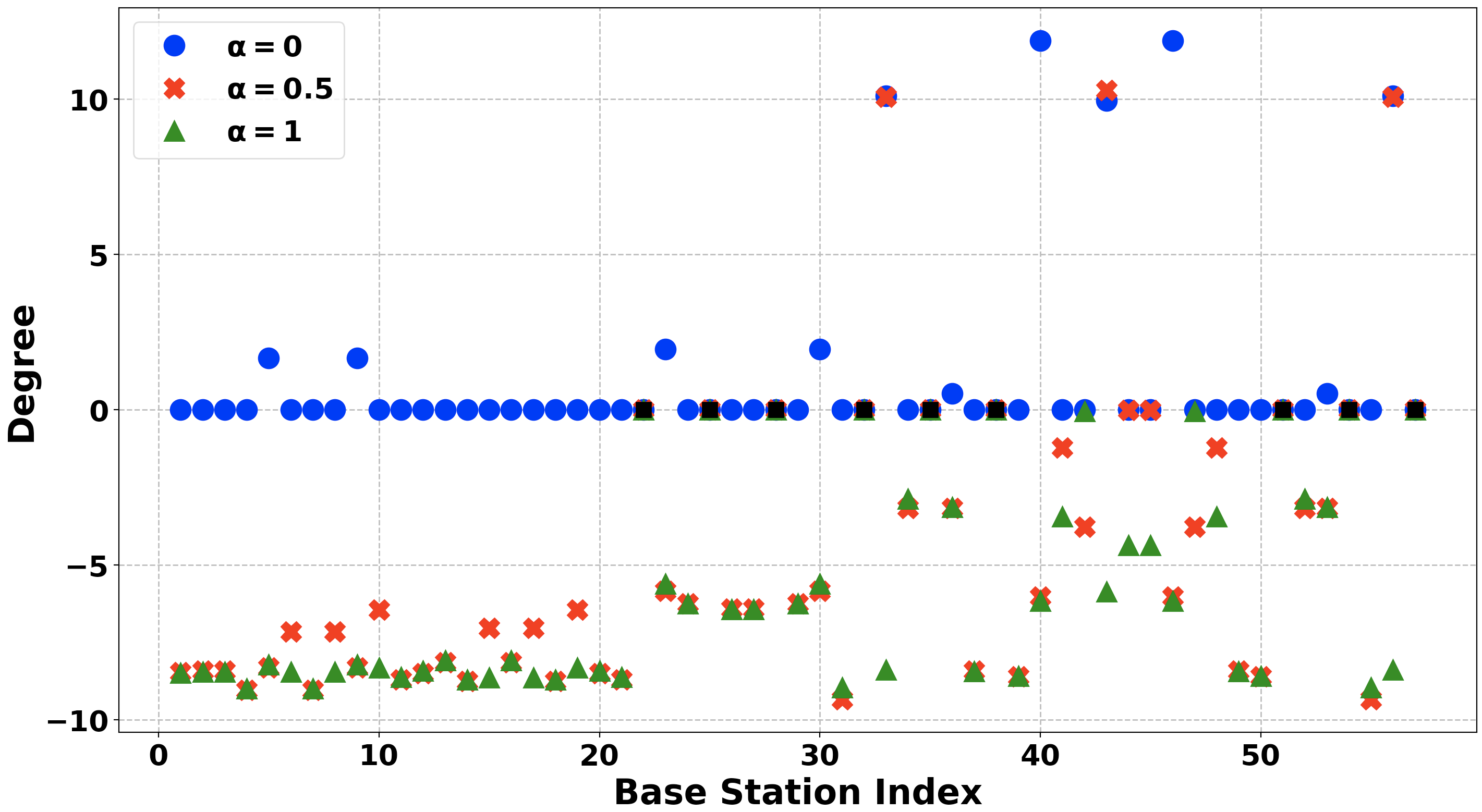}
\caption{{\small Optimized vertical tilts $\theta_i^*$ for: GUEs only ($\alpha=1$, green triangle), UAVs only ($\alpha=0$, blue circle), and both ($\alpha=0.5$, red cross). Black squares indicate cells that do not contribute to coverage.}}
\label{Fig:theta}
\end{figure}

Fig.~\ref{Fig:theta} shows the optimal values of the vertical electrical antenna tilts $\theta_i^*$ for each cell in the three cases $\alpha=\{1, 0, 0.5\}$. As expected, $\alpha=1$ (green triangle) entails optimizing all antenna tilts for only legacy ground users, and thus results in downtilted BSs. Conversely, $\alpha=0$ (blue circle) only caters for the four UAV corridors, and thus leads to uptilted BSs. As shown in Fig. ~\ref{Fig:theta}, not all BSs effectively contribute to optimizing the performance function, resulting in some vertical tilts remaining at the initial value of zero. In particular, BSs $22$, $25$, $28$, $32$, $35$, $38$, $51$, $54$, and $57$, that are shown by black squares in Fig. \ref{Fig:theta}, do not contribute to the performance function in any of the three simulated scenarios of $\alpha=0,\ 0.5$, and $1$. Lastly, $\alpha=0.5$ (red cross) seeks a coverage tradeoff between the ground and the UAV corridors, hence resulting in a small subset of BSs being uptilted, with the rest remaining downtilted.

\begin{figure}
     \centering
     \begin{subfigure}{0.87\columnwidth}
         \centering
 	 	  \includegraphics[width=0.99\columnwidth]{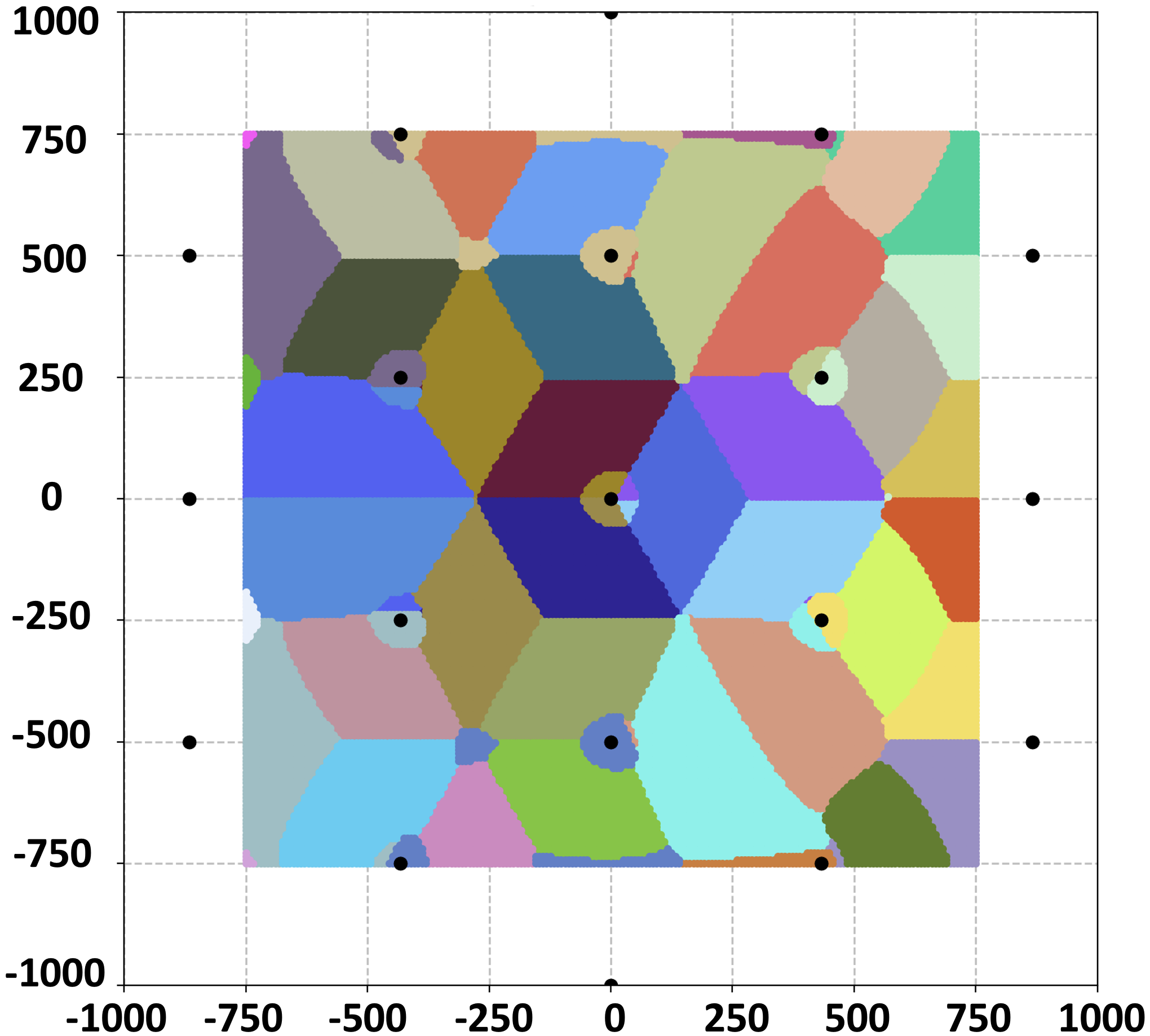}
         \caption{Resulting cell partitioning for ground users.}
        \vspace*{0.2cm} \label{Fig:cellPartitioningGUE}
     \end{subfigure}
     \begin{subfigure}{0.87\columnwidth}
         \centering
 	 	  \includegraphics[width=0.99\columnwidth]{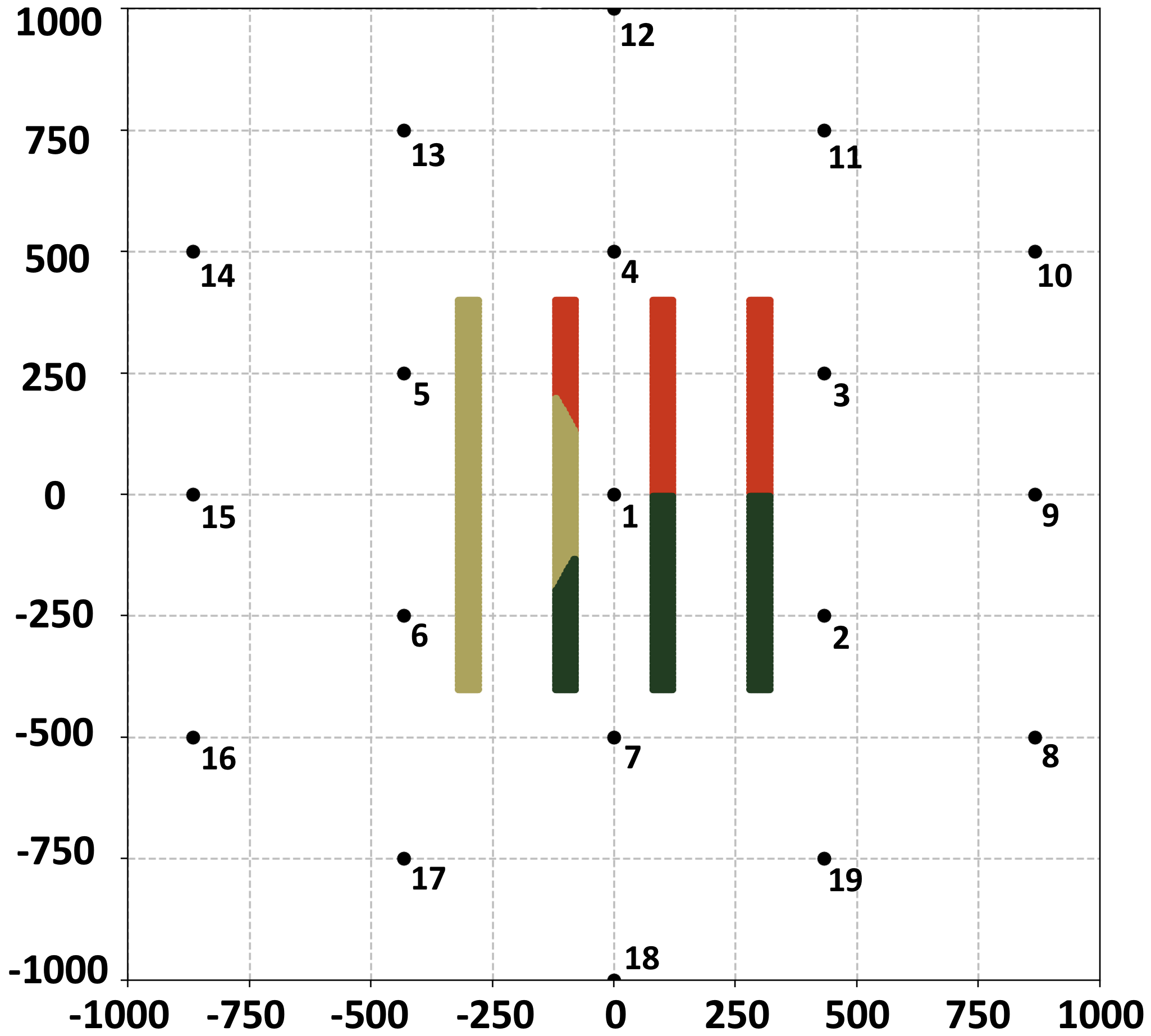}
 	 	  \caption{Resulting cell partitioning for UAV corridors.}
 		  \label{Fig:cellPartitioningUAV}
     \end{subfigure}
     \caption{Cell partitioning for (a) ground users and (b) UAVs when the vertical tilts are optimized for both (Fig.~\ref{Fig:theta}, $\alpha=0.5$). 
     }
     \label{Fig:cellPartitioning}
\end{figure}

Fig.~\ref{Fig:cellPartitioningGUE} and Fig.~\ref{Fig:cellPartitioningUAV} display the cell partitioning for ground users and UAV corridors, respectively, when the vertical tilts $\theta_i^*$ are optimized for both populations of end-devices (see case $\alpha=0.5$ in Fig.~\ref{Fig:theta}).
The figures show that the optimal tilt arrangement results in BSs $11$, $15$, and $19$ with respective azimuth orientations of $240^{\circ}$, $0^{\circ}$, and $120^{\circ}$, being devoted to covering UAV corridors, with the rest remaining downtilted. 
The resulting optimal cell partitioning is highly uncustomary and differs from a conventional hexagonal pattern.

\begin{figure}
\centering
\includegraphics[width=0.999\columnwidth]{
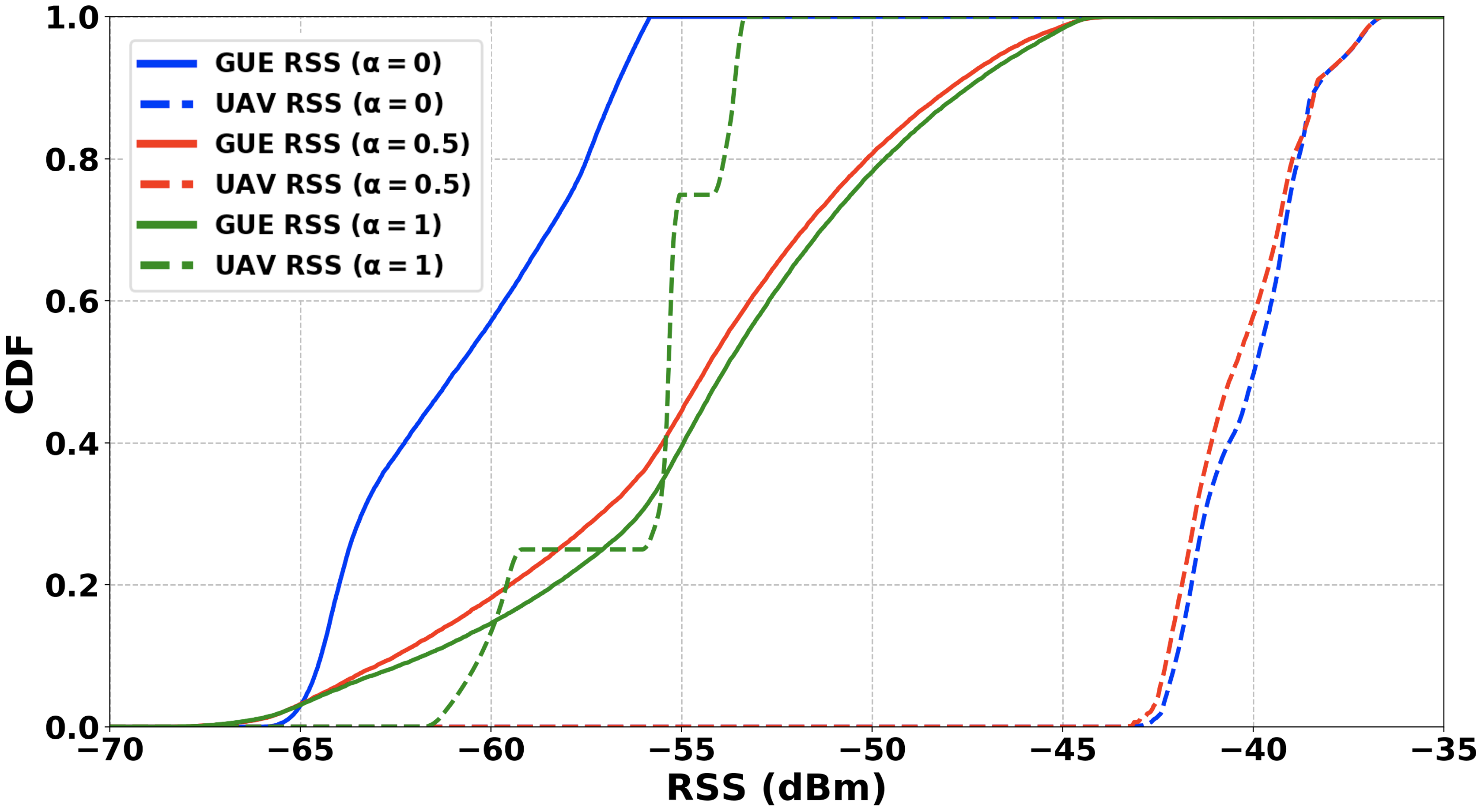}
\caption{{\small CDF of the RSS (dBm) at UAVs (dash-dash) and GUEs (solid) when the network is optimized for GUEs only ($\alpha=1$), UAVs only ($\alpha=0$), and both ($\alpha=0.5$).}}
\label{Fig:RSS}
\end{figure}

Finally, Fig.~\ref{Fig:RSS} shows the cumulative distribution function (CDF) of the RSS perceived by ground users (solid line) and UAVs (dash-dash line) when the network is optimized for ground users only ($\alpha=1$, green), UAVs only ($\alpha=0$, blue), and both ($\alpha=0.5$, red). Note that the ground user performance for $\alpha=1$ (green solid line) and the UAV performance for $\alpha=0$ (blue dash-dash line) can be regarded as respective upper bounds (in mean) since they entail optimizing all vertical tilts for ground users only and for UAVs only, respectively. Conversely, the ground user performance for $\alpha=0$ (blue solid line) and the UAV performance for $\alpha=1$ (green dash-dash line)\footnote{Note that the green dash-dash curve exhibits a staircase behavior, explained as follows. For $\alpha=1$, since tilts are optimized for GUEs only, no cell is pointing its antennas upwards (Fig.~\ref{Fig:theta}). UAVs are then just reached by the antenna sidelobe of the respective serving cells, with three cells in total serving all UAVs. All UAVs served by the same cell then experience very similar values of RSS, which are however different for each of the three cells.} can be regarded as respective baselines, obtained when the vertical tilts are chosen ignoring ground users and UAVs, respectively.
Fig.~\ref{Fig:RSS} shows that for $\alpha=0.5$ the proposed framework reaches a satisfactory tradeoff by: (i) significantly boosting the RSS at UAVs (red dash-dash line) compared to the baseline (green dash-dash line) and approaching the upper bound (blue dash-dash line), and (ii) nearly preserving the RSS at ground users (red solid line) compared to the upper bound (green solid line). While their evaluation falls beyond the scope of this work, both (i) and (ii) may have remarkable positive implications in terms of power control and interference mitigation, achievable rates, and even mobility management \cite{GerGarAza2022}.

\section{Conclusion}

In this article, we introduced a new mathematical framework for the analysis and design of UAV corridors in cellular networks, while considering a realistic network deployment, antenna radiation pattern, and propagation channel model. Our framework, based on quantization theory, allows to optimize the electrical tilts of existing ground cellular base stations to cater for both legacy ground users and UAVs flying along specified aerial corridors. 
Our case study showed that the ensuing electrical tilt arrangement may result in a highly non-obvious cell partitioning for both ground and UAV users, and that it can boost coverage along UAV corridors without degrading the perceived signal strength on the ground.

Our work is amenable to extensions from at least three standpoints:
(i) Rather than focusing on RSS, a proxy for coverage, a similar approach can be taken to optimize for signal-to-interference-plus-noise ratio (SINR). While we found the ensuing analysis to be tractable and insightful, it involves longer mathematical derivations and has been omitted from the present paper due to lack of space; (ii) Our case study assumed LoS condition on all links and a specific antenna arrangement. This could be modified to account for variable (either deterministic or probabilistic) LoS link conditions and to account for other radiation patterns, e.g., with beamformed synchronization signal blocks for initial access; and
(iii) Instead of optimizing the antenna tilts for a given cellular deployment, our mathematical framework could be repurposed to optimize the locations of the BSs themselves, to identify suitable sites for dedicated uptilted deployments or both.

\bibliographystyle{IEEEtran}
\bibliography{main}

\end{document}